\documentclass[runningheads]{cl2emult}
\begin{document}

\title*{The First Stars: Final Remarks}
\toctitle{The First Stars: Final Remarks}
\titlerunning{Final Remarks}

\author{Richard B. Larson}
\authorrunning{Richard B. Larson}
\institute{Department of Astronomy, Yale University, New Haven,
           CT 06520-8101, USA}
\maketitle
\vskip 1cm

   How did star formation begin in the universe?  Can we make any
credible predictions, and can we find any traces of the first stars?
We have only recently begun to be able to address these questions in
any depth, thanks to a rapidly developing understanding of the history
of the universe and to increasingly powerful instruments, and our
efforts to answer them have formed the subject of this stimulating first
conference on ``The First Stars''.  In these final remarks, I shall try
to summarize from a theorist's perspective some of the questions that
have been addressed here, and offer some brief comments on what we may
 have learned so far.

   {\bf How did the first stars form?}  In the standard hot-big-bang
picture, essentially no heavy elements were produced during the big
bang, and the first stars must therefore have formed without any heavy
elements.  This changes and simplifies the physics of star formation
compared with the present situation, since the important physical
processes then involve only various forms of hydrogen; the cooling of
the first star-forming clouds, for example, is controlled by molecular
hydrogen.  The thermal behavior of the first collapsing clouds is
becoming relatively well understood, thanks to the work of several
groups as reported here, and a general conclusion is that the first
star-forming clouds must have been hotter than present-day molecular
clouds by one or two orders of magnitude in temperature.  This higher
temperature means that thermal pressure must have been more important
for early star formation than it is at the present time, and also that
the Jeans mass must have been higher for similar cloud pressures or
densities.  Even in non-standard cosmologies, the amount of heavy
elements that can be produced during the big bang is still very small,
and probably too small to change the conclusion that heavy elements
played no important role in the formation of the first stars.

   {\bf When and where did they form?}  Recent progress in cosmology has
narrowed the class of popular models to several variants of the standard
CDM model which predict that the first collapsed structures formed at
redshifts between about 20 and 50 and had masses between about $10^5$
and $10^8\,$M$_\odot$.  The work reported here has mostly focused on a
`typical' case in which the first $3\sigma$ density peaks collapsed at
a redshift of $\sim\,$30 and formed bound structures with masses of the
order of $10^6\,$M$_\odot$.  In this case, the first population~III
stars are predicted to have formed at a redshift of $\sim\,$30 in small
systems whose total masses were of the order of $10^6\,$M$_\odot$ and
whose baryonic masses were of the order of $10^5\,$M$_\odot$.  The
formation of metal-free stars need not have occurred only at such high
redshifts, however; lower-amplitude primordial density peaks still
unenriched in heavy elements could have continued to collapse and form
metal-free stars at smaller redshifts, and the formation of metal-free
stars could conceivably have continued until much more recent times in
the least dense and most slowly evolving `backwaters' of the universe.
  
   {\bf What were their typical masses?}  In the detailed simulations
reported here, the dark matter in the first collapsing structures
virializes to form small dark halos, while the baryonic matter settles
into flattened configurations in which the Jeans mass is of the order of
$10^3\,$M$_\odot$.  This result does not seem to depend very much on the
details of the simulations, but only a few cases have been studied so
far.  Studies of present-day star formation suggest that the Jeans mass
may play an important role in determining typical stellar masses, and
the simulations reported here also suggest this, showing the formation
of a small number of dense collapsing gas clumps whose typical mass is
of the order of $10^3\,$M$_\odot$.  A considerable mass range for the
clumps is also suggested, extending from less than $10^2\,$M$_\odot$ to
more than $10^4\,$M$_\odot$.  Little tendency is found for these clumps
to fragment into smaller objects as the simulations are pushed to higher
densities, in agreement with our understanding of present-day star
formation which suggests that the outcome of the collapse of Jeans-mass
clumps is the formation of at most a small multiple system, so it seems
likely that the first stars were indeed typically very massive.

   {\bf Were any low-mass population~III stars formed?}  Fragmentation
to much smaller masses can occur if some of the gas collapses into thin
filaments, as indeed happens in some simulations.  The fragmentation of
such filaments is ultimately limited by the onset of high opacity to
the cooling radiation, and in metal-free gas the lower mass limit set
by opacity is comparable to the Chandrasekhar mass and somewhat above
one solar mass.  This is an important result because it means that no
metal-free stars should remain visible today; all such stars should by
now have evolved.  This may explain why we now see no metal-free stars,
although it is not yet clear whether we can argue the inverse, namely
that the fact that we see no metal-free stars means that the first stars
formed must have been exclusively massive.  Similar effects may explain
why we apparently see fewer than expected extremely metal-poor stars,
and it will be interesting to study the effect of a finite but low
metallicity on star formation to see whether there is a threshold
metallicity above which significant numbers of low-mass stars can form.

   {\bf What effects did they have?}  The apparent reionization of
the universe at a redshift larger than 5 could in principle have been
caused by a small number of massive population~III stars formed at high
redshifts, but the ionization history of the universe was probably more
complex than this, and the effects of the first stars may initially have
been rather local and limited by negative feedback effects.  One such
feedback effect might have been the dissociation of hydrogen molecules
by UV radiation before most of the gas became ionized; this might have
suppressed further star formation by removing the possibility of cooling
by molecular hydrogen.  Star formation might have picked up again when
larger regions of the universe collapsed and created systems with more
internal structure and densities high enough to provide self-shielding
from the dissociating UV radiation.  Simple predictions of ionization
effects are not possible in this more complex situation, but a general
expectation illustrated by numerical simulations is that the low-density
parts of the universe became ionized first and that the denser regions
took longer to become ionized.

   {\bf How did metal enrichment begin?}  The first stars must also
have produced the first heavy elements.  Stars more massive than about
200$\,$M$_\odot$ are predicted to collapse completely to black holes,
and therefore they should not contribute to heavy-element production.
But stars with masses between about 100 and 200$\,$M$_\odot$ are
predicted to disrupt entirely due to the pair-production instability,
so stars in this mass range are plausible candidates for the first
sources of heavy elements.  Stars with smaller masses, perhaps between
40 and 100$\,$M$_\odot$, may again collapse to black holes and not
contribute to nucleosynthesis, but still smaller stars with masses
between 10 and 40$\,$M$_\odot$ may produce type~II supernovae and
provide a second source of heavy elements, if significant numbers
of such stars were formed.  It is less straightforward to understand
how the heavy elements produced by these stars became mixed into the
surrounding medium and incorporated into subsequent generations of
stars; not until this had happened could the formation of stars of
finite metallicity begin.  The dispersal and mixing of heavy elements
is a complex process, and it was almost certainly not as efficient as
has usually been assumed in simple models of the chemical evolution
of the universe; chemical enrichment may initially have been quite
localized.  Supernova-driven winds and galaxy mergers may have
contributed to dispersing the heavy elements, but the universe may well
have remained chemically very inhomogeneous up to the present time.

   {\bf Where are the first stars or their products now?}  If the first
stars typically had masses of the order of $10^3\,$M$_\odot$, they
would mostly have collapsed into black holes with masses of this order.
Simulations that keep track of the locations of the first stars or their
remnants show that these first condensed objects, which formed in the
densest parts of the universe, typically became incorporated through
successive mergers in systems of larger and larger size, and typically
ended up in the inner parts of present-day large galaxies.  If
significant numbers of black holes with masses of $10^3$ or even
$10^4\,$M$_\odot$ were present at early times in the inner parts of
large galaxies, they might have played a role in the formation of the
central supermassive black holes of AGNs.  One of the remnants of the
early population~III stars might have served as a seed for building up a
supermassive black hole by accretion, or many of them might have merged
into a single much larger black hole because of strong gravitational
drag effects in the dense environment of a forming galactic nucleus.
Conceivably, most of the first stars or their remnants could have ended
up in the central black holes of AGNs!

   The heavy elements produced by the first stars may also have
ended up mostly in the inner parts of large galaxies.  Heavy-element
abundances in galaxies are observed to increase towards their centers,
and also to increase systematically with galactic mass.  Neither of
these trends is fully explained by standard models, but both might be
explainable if heavy elements produced by the first stars contributed
significantly to the observed abundances.  The innermost parts of
the largest galaxies were the first regions to make stars and heavy
elements, and if the first stars formed were predominantly massive,
high abundances of heavy elements could have been produced in these
regions.  The high metallicities of some quasars might also be
explainable in this way.

   {\bf Were they closely related to the oldest observed stars?}
The hypothetical $10^6$-M$_\odot$ systems that formed the first stars
probably cannot be identified with any observed systems, and they
probably did not form the oldest observed stars because they would have
been too short-lived and too weakly bound to retain any gas or heavy
elements.  The first systems capable of self-enrichment were probably
larger systems formed somewhat later by the collapse of larger-scale
cosmological structures.  These early star-forming systems may have
resembled dwarf galaxies, but they probably cannot be identified with
observed dwarf galaxies like those in the Local Group since the latter
objects inhabit low-density regions and are actually relatively young
systems, being dominated in many cases by stars of intermediate age.
Thus the observed dwarfs may have been `stragglers' that formed
relatively late in low-density regions of the universe, and not the
birthplaces of the first stars.  Globular clusters, another once-popular
candidate for the sites of the first star formation, are almost
certainly not primordial self-enriched objects, since their internal
chemical homogeneity cannot be explained without very contrived
assumptions unless they were formed in larger pre-existing systems
that provided an environment for chemical enrichment and mixing to
take place.

   {\bf How did the first observed stars form?}  A notable property of
the dwarf galaxies in the Local Group is that they appear to have a
minimum mass of about $2 \times 10^7$~M$_\odot$, regardless of how faint
or metal-poor they may be.  This may be the minimum mass that a galaxy
needs to retain gas and heavy elements, and thus to allow continuing
star formation and self-enrichment to occur.  In fact, this mass is
about the minimum required for a galaxy to bind ionized gas at a
temperature of $10^4\,$K; retaining ionized gas is necessary to sustain
star formation and chemical enrichment because massive stars quickly
ionize the surrounding medium as soon as they form, and the interstellar
gas in a typical galaxy goes through many cycles of ionization and
recombination before being incorporated into stars.  This cycling
process also plays an important role in chemical enrichment, since the
heavy elements produced by supernovae are probably dispersed and mixed
mainly in an ionized medium.  Many complex astrophysical processes must
therefore have occurred prior to the formation of the first observed
stars, and star-forming systems at least as massive as present-day dwarf
galaxies may have been required.  These early star-forming systems might
also have been the birthplaces of the first globular clusters.  However,
nearly all of them would by now have been destroyed by being merged into
larger galaxies, and the globular clusters may be their only surviving
remnants.

   {\bf What can we learn from element abundances?}  Much attention has
been given at this meeting to chemical abundances in old stars, but
this subject has many intricacies, and their  implications for our
understanding of early star formation and galaxy evolution are not yet
very clear.  Some of the abundance patterns in very metal-poor stars
seem compatible with element synthesis in predominantly massive stars,
and this might be consistent with enrichment by an initial metal-free
population consisting mostly of massive stars; however, the origin of
the heavy elements in these very metal-poor stars cannot yet be
identified with certainty.

   We are just beginning, in any case, to appreciate the true complexity
of the chemical enrichment of galaxies and the universe.  Clearly it is
completely misleading to imagine that heavy-element abundances are
correlated in any simple way with time and can be used as a clock;
instead, it is clear that the densest parts of the universe and of
individual galaxies evolved more rapidly and became chemically enriched
much earlier than the less dense regions.  The universe must therefore
have evolved in a chemically highly inhomogeneous way; ``old'' and
``metal-poor'' are not synonymous.  Even in the solar neighborhood in
our own Galaxy, chemical enrichment must have been a very non-uniform
process, since the metallicities of nearby stars and clusters show a
large scatter and only a weak trend, if any, with age.  All of the
standard models of chemical evolution fail badly to account for these
observations, and we need to go back to the drawing board with these
models because our present understanding of this subject is still
primitive.

   So, much has been learned, but much remains to be done.  It is an
encouraging sign of progress that we have even been able to make a start
in answering some of the questions mentioned above.  Let us look forward
to many more fruitful meetings as we continue the quest to understand
the first stars.

\end{document}